\documentclass[aps,prb,twocolumn,superscriptaddress,amsmath,amssymb]{revtex4}

\usepackage{amssymb}
\usepackage{graphicx}
\usepackage{tabularx}
\usepackage{bm}
\usepackage{epsfig}
\usepackage[ansinew]{inputenc}

\begin{document}

\title{Influence of spin glass-like magnetic relaxation on the zero-field-cooled exchange bias effect}

\author{L. T. Coutrim}
\affiliation{Instituto de F\'{\i}sica, Universidade Federal de Goi\'{a}s, 74001-970, Goi\^{a}nia, GO, Brazil}

\author{E. M. Bittar}
\affiliation{Centro Brasileiro de Pesquisas F\'{\i}sicas, 22290-180, Rio de Janeiro, RJ, Brazil}

\author{F. Garcia}
\affiliation{Centro Brasileiro de Pesquisas F\'{\i}sicas, 22290-180, Rio de Janeiro, RJ, Brazil}

\author{L. Bufai\c{c}al}
\email{lbufaical@ufg.br}
\affiliation{Instituto de F\'{\i}sica, Universidade Federal de Goi\'{a}s, 74001-970, Goi\^{a}nia, GO, Brazil}

\date{\today}

\begin{abstract}
The zero-field-cooled exchange bias (ZEB) effect is a remarkable phenomenon recently reported for some reentrant spin glass-like compounds. In this work, the time-evolution of magnetization is thoroughly investigated for two ZEB materials in order to figure out the role played by the spin glass-like phase on such effect. La$_{1.5}$Sr$_{0.5}$CoMnO$_{6}$ and La$_{1.5}$Ca$_{0.5}$CoMnO$_{6}$ were chosen as representative samples of ZEB systems, since the former compound presents the largest ZEB reported so far, while the second has a much smaller effect, despite being structurally/chemically similar. Comprehensive magnetic measurements were carried on both samples, and the results are discussed in terms of the amount and time-evolution of the spin glass-like phase under the influence of a varying field. We also propose a phenomenological model, based on the pinning of spin glass-like moments and on the dynamics of their magnetic relaxation, to explain the asymmetry observed in the hysteresis loops. The good agreement between the simulated and experimental results confirms our hypothesis that the spin glass-like phase is key to the ZEB effect.
\end{abstract}

%\pacs{75.50.Lk, 75.30.Gw, 75.60.Jk, 75.47.Lx}

\maketitle

\section{Introduction}

Spin glass (SG) materials are significantly different from most condensed matter systems, such as conventional ferromagnets, liquid crystals, superconductors, etc. The fundamental difference is that, for the previous listed systems, there are well-known symmetries that allow mathematical simplifications and physical insights, admitting easier physical modeling \cite{Fischer}. Conversely, SG present quenched magnetic disorder, for which there is no evident long-range order, leading to non-obvious phase transitions and broken symmetries \cite{Stein}. Therefore, the intriguing properties of SG, especially its dynamics, are not well understood yet \cite{Fischer,Stein,Mydosh}.

Another subject of great academic and technological interest is the exchange bias (EB) effect, characterized by a shift of the magnetic hysteresis loop [$M(H)$] along the magnetic field ($H$) axis. Usually, this effect is ascribed to the unidirectional exchange anisotropy formed at the interfaces of antiferromagnetic (AFM) and ferromagnetic(FM)/ferrimagnetic(FIM) phases in heterogeneous systems, being conventionally observed after the system is cooled through its N\'{e}el temperature under an applied magnetic field \cite{Nogues}. This conventional EB (CEB) effect is a well-known phenomenon discovered around 60 years ago, and improvements in techniques for the production of heterostructures have renewed its interest, enabling the manufacture of multifunctional devices using strongly correlated electronic materials \cite{Dagotto}.

SG and EB merge in recently discovered materials that display a shift of their hysteresis loop even when they are cooled from an unmagnetized state down to low temperature ($T$) in zero field. This is commonly referred to as zero field cooled exchange bias (ZEB) effect. Although different scenarios were proposed to explain such effect on distinct systems, all ZEB materials reported so far have the reentrant spin glass-like (RSG) behavior as a common feature, which is characterized by an SG-like state concomitant with conventional magnetic phases \cite{Mydosh,Wang,Maity,Nayak,Tian}. Due to their intrinsic inhomogeneity, double-perovskite (DP) compounds usually present structural and magnetic disorder \cite{Vasala,Serrate}, being thus prospective candidates to exhibit RSG behavior and ZEB effect. Indeed, the majority of observed ZEB materials present perovskite structure \cite{Maity,MeuPRB,Murthy,MeuJMMM,Huang,Xie}.

The La$_{1.5}$Sr$_{0.5}$CoMnO$_{6}$ (LSCMO) compound stands out as having the largest ZEB effect reported so far \cite{Murthy}. In addition, it was observed that replacing Sr by Ca [La$_{1.5}$Ca$_{0.5}$CoMnO$_{6}$ (LCCMO)], also gives rise to an RSG material presenting ZEB effect, however its $M(H)$ loop-shift is  one order of magnitude smaller than that of the Sr-based compound \cite{MeuJMMM}. This difference can be understood, because structural, electronic and magnetic properties are strongly intercorrelated on DP systems \cite{Vasala,Serrate}, small differences in any of these properties could critically affect the ZEB effect.

Although, the RSG behavior is a common feature of all ZEB materials, the microscopic mechanisms responsible for such effect are not fully understood. In this respect, the present work studies two DPs (LSCMO and LCCMO), which are representative examples of RSG and ZEB materials. Comprehensive magnetization measurements as a function of applied magnetic field and time [$M(H,t)$] were carried out, in order to shed light on the role played by the glassy magnetic phases on the ZEB effect. For comparison, we have also investigated the La$_{2}$CoMnO$_{6}$ (LCMO) DP, which is a non-RSG and non-ZEB material \cite{Singh,Burnus,Mir}. Our results clearly indicate that the coupling between the SG-like, FM, and AFM phases is paramount for the appearance of the ZEB effect. The different magnetic properties observed for the RSG DP compounds are discussed mainly in terms of the magnetic relaxation of the SG-like phase dynamics, which we describe by a proposed phenomenological model. This model predicts the horizontal shift observed in the $M(H)$ loops, with good agreement between experimental and calculated results, indicating that the time-evolution of the SG-like magnetization plays an important role in the ZEB effect.

\section{Experimental details}

Polycrystalline samples of LSCMO, LCCMO and LCMO were synthesized by conventional solid state reaction, as described in the Supplementary Material (SM) \cite{SM}. X-ray powder diffraction patterns revealed the formation of single phase DP structures for all compounds. The Rietveld refinement have indicated that LCMO and LCCMO grow in monoclinic $P2_1/n$ space group \cite{MeuJMMM}, while LSCMO forms in rhombohedral $R$-$3c$ space group, in agreement with previous reports \cite{Murthy}.

Magnetic measurements were carried out using a Quantum Design PPMS-VSM magnetometer. All $M(H)$ loops were performed at $T=5$ K up to a maximum magnetic field of $H_{max}=\pm90$ kOe. Since both the SG-like behavior and the EB effect are extremely sensitive to the cooling process \cite{Gokemeijer}, we followed the same protocol for each measurement performed on all investigated samples. For each measurement, the system was slowly zero field cooled (ZFC) from the paramagnetic state down to $T=5$ K, followed by a 10 minutes wait time, to guarantee the thermal stabilization. From one measurement to another the sample was warmed up to the paramagnetic state and the coil was demagnetized in the oscillating mode, in order to prevent the presence of trapped current on the magnet and ensure a reliable ZFC process.

\section{Results and analysis}

\subsection{Experimental results}

\begin{figure}
\begin{center}
\includegraphics[width=0.5 \textwidth]{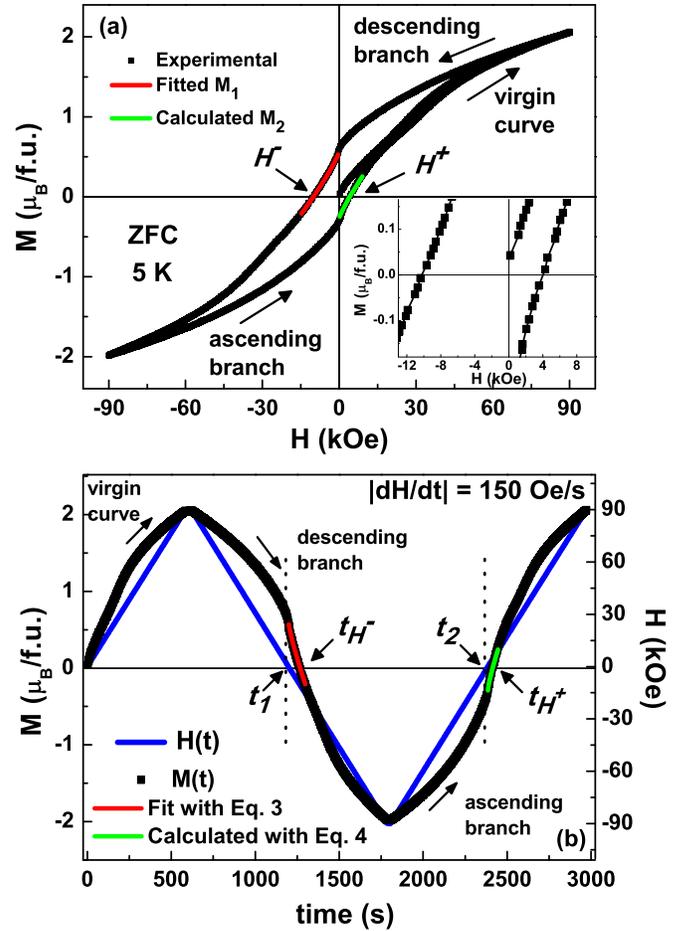}
\end{center}
\caption{(a) $M(H)$ loop of LSCMO at $T=5$ K and $|dH/dt| = 150$ Oe/s. Red and green solid lines are the calculated $M_{1}$ (Eq. \ref{Eq3}) and $M_{2}$ (Eq. \ref{Eq4}) stretches, respectively. Inset shows zoom in around $M=0$, evidencing the shift along $H$-axis. (b) The same hysteresis loop for LSCMO, now displayed in $M(t)$ mode. The blue solid line is the magnetic field time dependence.}
\label{FigMxH_Sr}
\end{figure}

The magnetization ($M$) of SG-like systems exhibits an unusual time dependence, which can lead to intriguing phenomena such as aging, rejuvenation and memory effects \cite{Fischer,Stein,Mydosh}. Hence, it is expected that $M(H)$ loops measured with different $H$ sweep rate ($dH/dt$) might lead to distinct $M$-relaxation of the SG-like phase, which consequently affects the ZEB effect. We measured $M(H)$ loops for both LSCMO and LCCMO samples in ZFC mode, with $|dH/dt|=150$ Oe/s, see Fig. \ref{FigMxH_Sr}(a) for the LSCMO compound. The measurement shows a clearly closed $M(H)$ loop, symmetric in respect to the $M$-axis and shifted to the left along the $H$-axis, as evidenced in the inset. These $M(H)$ loops are a superposition of three different magnetic phase contributions, namely, the hysteresis, due to the FM and SG-like phases, overlapped with a linear $H$-dependent contribution given by the AFM phase \cite{Mydosh,Murthy}. The shift along the $H$-axis is a measure of the EB field, herein defined as $H_{EB}=|H^{+}+H^{-}|/2$, where $H^{+}$ and $H^{-}$ are the positive and negative coercive fields, respectively. For LSCMO we obtained $H_{EB}^{exp}=3127$ Oe, while LCCMO showed $H_{EB}^{exp}=165$ Oe (see SM \cite{SM}). The $H$ values are herein approximated to integer numbers, and the experimental errors are neglected. Since $M$ depends on $H$, which in turn depends on time ($t$), Fig. \ref{FigMxH_Sr}(a) can be displayed as a function of time [$M(t)$], see Fig. \ref{FigMxH_Sr}(b), which is suitable to our purpose of understanding how the dynamics of the SG-like phase acts on the ZEB effect.

An evidence that the ZEB effect is critically affected by SG-like phase is the fact that $H_{EB}$ changes significantly depending on $|dH/dt|$. For LSCMO, a $M(H)$ loop with $|dH/dt| = 100$ Oe/s, leads to $H_{EB}^{exp} = 3100$ Oe, lower than $H_{EB}^{exp} = 3127$ Oe obtained with $|dH/dt| = 150$ Oe/s. For $|dH/dt| = 50$ Oe/s, the shift decreases even further to $H_{EB}^{exp} = 2980$ Oe. Interestingly, for LCCMO the opposite trend was observed. $M(H)$ loops with $|dH/dt| = 50$, 100 and 150 Oe/s resulted in $H_{EB}^{exp} = 165$, 173 and 185 Oe, respectively. See table \ref{T1}. 

Glassy magnetic systems present a very long-lasting $M$-relaxation. Therefore, to attest the influence of the SG-like phase on the ZEB effect, we pause the $M(H)$ loop at a given magnetic field value for a time-interval, leading to the relaxation of the magnetization, and subsequently completing the loop. 

\begin{figure}
\begin{center}
\includegraphics[width=0.5 \textwidth]{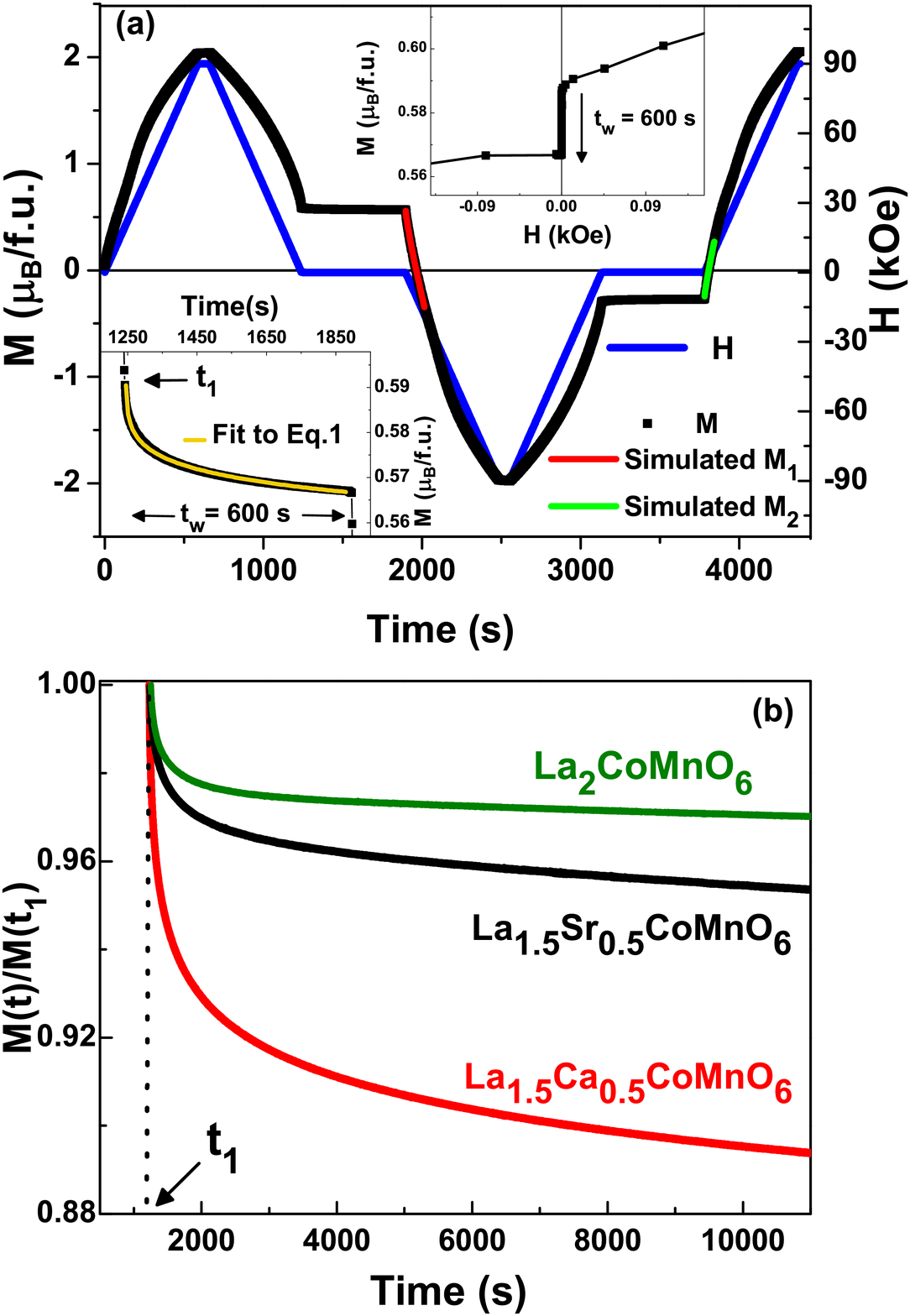}
\end{center}
\caption{(a) $M(t)$ curve of LSCMO ($T=5$ K, $|dH/dt| = 150$ Oe/s), paused at $H = 0$ for a time-interval $t_{w} = 600$ s, in the ascending and descending branches. Red and green lines are $M_{1}$ (Eq. \ref{Eq5}) and $M_{2}$ (Eq. \ref{Eq6}) stretches, respectively. The blue solid line is the magnetic field time dependence. The insets evidence the descending branch $M$-decay at $H = 0$ in both $M(H)$ (upper inset) and $M(t)$ (bottom inset) modes, and the yellow solid line shows the fitting of the $M$-decay at $H$ = 0 with Eq. \ref{Eq1}. (b) Normalized $M$-decay for $t_w = 10^{4}$ s for LSCMO and LCCMO RSG samples, and for LCMO non-RSG and non-ZEB sample.}
\label{Fig_Wait0T_Sr}
\end{figure}

Firstly we chose to pause $M(H)$ loops, with $|dH/dt| = 150$ Oe/s, at $H = 0$ on the descending branch, which is herein defined as isothermal remnant $M$-relaxation [$IRM(t)$] [lower inset on Fig. \ref{Fig_Wait0T_Sr}(a)] \cite{IRM}. Moreover, we also pause at $H = 0$ on the ascending branch, as depicted in Fig. \ref{Fig_Wait0T_Sr}(a). The wait time were the same for both branches, for time-intervals of $t_w=600$, 3600 and 10$^{4}$ s. In these cases, SG-like spins are expected to relax in a way to decrease the magnetization, thus reducing both $|H^{-}|$ and $H^{+}$. As will be discussed later, the effect of SG-like relaxation is stronger on the descending branch and consequently the decrease of $|H^{-}|$ is more pronounced. This results in a systematic reduction of $H_{EB}^{exp}$ as $t_w$ increases. The observed values for LSCMO $H_{EB}^{exp}$ were 3082, 3060 and 3045 Oe for $t_w=600$, 3600 and 10$^{4}$ s, respectively, as can be verified in Fig. \ref{Fig_Wait0T_Sr}(a) [displayed in $M(t)$ mode for $t_w = 600$ s]. The same trend was observed for LCCMO, \textit{i.e.}, the decrease of the ZEB effect as function of $t_w$, from $H_{EB}^{exp}=165$ Oe ($t_w=0$) to 157 Oe ($t_w=600$ s), 98 Oe ($t_w=3600$) and 83 Oe ($t_w=10^{4}$) (see SM \cite{SM}).

One might argue that the decrease of $H_{EB}^{exp}$ could be explained by changes in other than the SG-like phase. Nevertheless, a remarkable feature of SG-like system is that, even for the longest $t_w$, the magnetization is expected to decay continuously \cite{Fischer,Stein,Mydosh}. Then, in order to distinguish the $M$-decay behavior from that of a conventional ferromagnetic system, we also investigated the $t_w$ effect in the $M(H)$ loop of LCMO material. This FM insulator DP was extensively investigated due to its room-$T$ magnetodielectric properties, and it presents no RSG behavior, neither EB effect \cite{Singh,Burnus,Mir,Manna,Murthy3}. Fig. \ref{Fig_Wait0T_Sr}(b) shows the normalized $M$-decay [$M(t)/M(t_1)$] at the descending branch for $t_w = 10^4$ s curves for all samples. As can be noted, $M(t)/M(t_1)$ rapidly drops to a nearly constant value for the non-RSG sample, while it keeps continuously decreasing for the RSG samples, in a faster rate for the Ca-based material. The fact that $H_{EB}^{exp}$ continuously decreases as $t_w$ increases on LSCMO and LCCMO RSG samples, while for LCMO conventional magnetic material it was not observed any loop-shift (see SM\cite{SM}). These results are a strong evidence of the influence of the SG-like phase on the ZEB effect.

\begin{figure}
\begin{center}
\includegraphics[width=0.5 \textwidth]{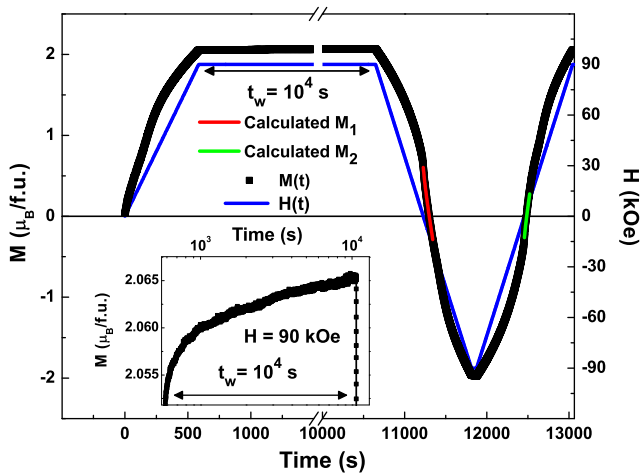}
\end{center}
\caption{$M(t)$ curve of LSCMO ($T=5$ K, $|dH/dt| = 150$ Oe/s), paused at $H_{max} = 90$ kOe for a time-interval $t_{w} = 10^{4}$ s. Red and green solid lines are the calculated $M_{1}$ (Eq. \ref{Eq3}) and $M_{2}$ (Eq. \ref{Eq4}) stretches, respectively. The inset evidences the $M$-relaxation at $H_{max} = 90$ kOe. The blue solid line is the magnetic field time dependence.}
\label{Fig_Wait9T_Sr}
\end{figure}

We now paused the $M(H)$ loop at the end of the virgin branch ($H_{max} = 90$ kOe), and subsequently completed the loop. Fig. \ref{Fig_Wait9T_Sr} displays the $M(t)$ curve obtained for $t_w = 10^{4}$ s on LSCMO, where the inset highlights the $M(t)$ relaxations at $H_{max} = 90$ kOe. The SG-like phase $M$-relaxation certainly affects the loop (on both descending and ascending branches), but since the loop was paused only at $H = 90$ kOe, $H^{-}$ is expected to vary more significantly than $H^{+}$, leading to the increase of $H_{EB}$. For LSCMO, $|dH/dt| = 150$ Oe/s and $t_w = 600$, 1800, 3600 and 10$^{4}$ s, $H_{EB}^{exp}$ is enhanced to 3198, 3326, 3396 and 3527 Oe, respectively. The same overall behavior was observed for LCCMO, for which the measurements with $t_w = 0$, 600, 1800, 3600 and 10$^{4}$ s resulted in $H_{EB}^{exp} = 165$, 188, 227, 270 and 292 Oe, respectively \cite{SM}.

In order to ensure that there are no extrinsic artifacts affecting the measurements, such as trapped current in the superconducting coils, we performed the same protocol described above, with $t_w$ = 10$^{4}$ s, for a Palladium standard, as well as, the non-ZEB LCMO samples. The $M(H)$ curves obtained for these materials are symmetric in relation to the horizontal and vertical axis, giving further evidence that the ZEB effect observed is related to the RSG behavior and not to trapped flux in the magnet, see SM \cite{SM}.

All these results obtained for both samples can be readily understood in terms of our phenomenological model, which is based on the pinning and magnetic relaxation of the SG-like moments. According to it, changes in $|dH/dt|$ affect the relaxation of the SG-like spins and the balance between the SG-like, FM and AFM phases present in the systems. As will be discussed below, these variations have a direct impact on $H^{+}$ and $H^{-}$, and consequently on $H_{EB}$.

\subsection{The ZEB model}

An usual EB effect is explained in terms of the coupling between two different magnetic phases, one of pinned moments and the other with reversing spins, which are being driven by the applied magnetic field. Similarly, our model for ZEB systems is based on the coupling between the reversing spins of the FM phase with the pinned spins of the SG-like phase. In a $M(H)$ loop of a ZEB material, for instance, in the descending branch for $H<0$, due to the relatively slow relaxation, not all spins of the SG-like phase will be flipped toward the negative field direction. Some of the moments will still point toward the positive $H_{max}$ field previously applied.

In Fig. \ref{FigMxH_Sr}(b), $t_1$ and $t_2$ represent the times when $H = 0$, while $t_{H^{-}}$ and $t_{H^{+}}$ correspond to the instants when $M = 0$ (in the descending and ascending branches, respectively). The magnetization at the $t$-interval $t_1 \leq t \leq t_{H^{-}}$ is defined as $M_1$ stretch, and the $t_2 \leq t \leq t_{H^{+}}$ interval as $M_2$ stretch. A key point of our phenomenological model is how the dynamics of the SG-like moments are affected by the magnetic history of the $M(H)$ loop. The amount $M_{SG}$ pinned on the opposite $H$-direction will change asymmetrically in $M_1$ and $M_2$ stretches, due to their magnetic history, \textit{i.e.}, while the $M_1$ stretch history is associated to the virgin curve and half of the descending branch, $M_2$ is correlated to the virgin branch, the whole descending and half of the ascending branches.

To compute how the SG-like phase may affect the hysteresis loop, we estimated the stretches of the $M(H)$ curves between $H=0$ and $M = 0$ (\textit{i.e.}, the regions encompassing $M_1$ and $M_2$ in the descending and ascending branches, respectively). To calculate both stretches, the following steps were followed: i) the magnetization's time-evolution of the SG-like phase was taken into account; ii) the time-evolution of the AFM and FM phases due to the applied magnetic field was also considered; iii) from the parameters obtained in steps i and ii, the magnetization of each stretch was calculated. For the ascending branch stretch, it was assumed, as an approximation, that nearly half of the SG-like moments had flipped toward negative $H$ direction, while the other half is relaxing but still pointed toward the opposite direction. From the resulting calculated curves, $H^{+}$ and $H^{-}$ could be computed, allowing the estimation of $H_{EB}$.

The first step considers the dynamics of the SG-like phase during the $M(H)$ loop measurement. The time-evolution of SG systems has been extensively debated in the last decades. Many models have been proposed to describe curves like the $IRM(t)$  [see Ref. \cite{IRM} and the lower inset of Fig.  \ref{Fig_Wait0T_Sr}(a)]. The Stretched Exponential Model is generally admitted to be the most relevant to account for these curves in conventional SG materials \cite{Fischer,Mydosh}. Given that we are not dealing here with canonical SG, but with RSG materials, which exhibit a non-negligible contribution to the magnetization from the FM phase, a term must be added to the Stretched Exponential equation in order to account for this contribution, yielding
\begin{equation}
M_{SG}(t)=M_{sp} + M_{0}e^{-\left[(t-t_{1})/t_{p}\right]^{n}}, \label{Eq1}
\end{equation}
where $M_{sp}$ corresponds to the spontaneous magnetization of the FM phase, $M_{0}$ is the initial magnetization of the SG-like phase at the instant $t_1$, finally, $t_{p}$, and $n$ ($0<n<1$) are the time and the time-stretch exponential constants, respectively. The parameters obtained from the fit of the experimental $IRM(t)$ curve with Eq. \ref{Eq1} are displayed on Table \ref{T1}.

\begin{table*}
\tiny
\centering
\caption{Fitted (Eqs. \ref{Eq1}, \ref{Eq2} and \ref{Eq3}) and calculated (Eq. \ref{Eq4}) parameters obtained from $IRM(t)$ and $M(H)$ measurements for LSCMO and LCCMO. The goodness-of-fit parameters $\chi^{2}_{1}$ and $\chi^{2}_{3}$ represent the reduced $\chi^{2}$ of the fittings with Eqs. \ref{Eq1} and \ref{Eq3}, respectively.}
\label{T1}
\begin{tabular}{c|c|c|c|c|c|c|c|c|c|c|c|c|c|c|c|c|c|c}
\hline \hline
Sample & $|dH/dt|$ & $M_{sp}$ & $M_{0}$ & $t_{p}$ & $n$ & $\chi^{2}_{1}$ & $A$ & $B$ & $r$ & $\chi^{2}_{3}$ & $t_{1}$ & $t_{2}$ & $t^{exp}_{H^{-}}$ & $t^{exp}_{H^{+}}$ & $t^{calc}_{H^{+}}$ & $H_{EB}^{exp}$ & $H_{EB}^{calc}$ \\
& (Oe/s) & ($\mu_B$/f.u.) & ($\mu_B$/f.u.) & (s) & & & ($\mu_B$/f.u.) & ($\mu_B$/f.u.) & & & (s) & (s) & (s) & (s) & (s) & (Oe) & (Oe) \\
\hline
& 50 & 0.266 & 0.361 & 1.908$\times$10$^{11}$ & 0.098 & 1.2$\times$10$^{-4}$ & 0 & 0.008 & 0.806 & 5.4$\times$10$^{-4}$ & 3524.1 & 7053.9 &  & 7113.7 & 7137.4 & 3072 & 2980 \\
LSCMO & 100 & 0.262 & 0.362 & 1.924$\times$10$^{11}$ & 0.096 & 2.4$\times$10$^{-4}$ & 0 & 0.013 & 0.814 & 1.8$\times$10$^{-5}$ & 1775.1 & 3554.5 & 1876.9 & 3594.5 & 3594.3 & 3090 & 3100 \\
& 150 & 0.258 & 0.358 & 3.634$\times$10$^{11}$ & 0.094 & 3.6$\times$10$^{-4}$ & 0 & 0.017 & 0.837 & 4.8$\times$10$^{-5}$ & 1197.6 & 2385.2 & 1266.3 & 2412.5 & 2412.2 & 3105 & 3127 \\
\hline
& 50 & 0.855 & 0.110 & 652.0 & 0.359 & 3.0$\times$10$^{-4}$ & 0.0029 & 3.4$\times$10$^{-4}$ & 1.373 & 3.8$\times$10$^{-6}$ & 3517.2 & 7045.8 & 3691 & 7212.2 & 7209.5 & 185 & 253 \\
LCCMO& 100 & 0.854 & 0.099 & 310.7 & 0.391 & 5.9$\times$10$^{-4}$ & -0.034 & 0.038 & 1.034 & 1.3$\times$10$^{-5}$ & 1792.8 & 3572.4 & 1881.4 & 3657.7 & 3657.1 & 173 & 195 \\
& 150 & 0.854 & 0.099 & 236.6 & 0.398 & 8.8$\times$10$^{-4}$ & -0.027 & 0.034 & 1.052 & 1.4$\times$10$^{-5}$ & 1195.8 & 2383 & 1254.7 & 2440 & 2439.9 & 165 & 173 \\
\hline \hline
\end{tabular}
\end{table*}

As stated above, to compute the $M_1$ and $M_2$ stretches, the effect of $|dH/dt|$ on the FM and AFM phases must be taken into account. Several functional forms have been proposed to fit the hysteresis curves of magnetic materials and, in general, each compound is better described by a particular model, \textit{i.e.}, there is not a universal equation that models $M(H)$ loops \cite{Trutt}. Since our purpose is not to understand microscopically how the magnetic field affects the AFM and FM phases, but only to fit the contribution from these phases to $M_1$ and $M_2$, our choice was for the simplest model. Thus, the following equation is suitable for our purpose:
\begin{equation}
M_m(H)= A'H + B'H^{r}. \label{Eq2H}
\end{equation}
In this equation, $A'$ is related to the linear $H$ dependence of the AFM phase, and $B'$ and $r$ account for the non-linear contribution of the FM phase to magnetization.

Alternatively, since the field sweep rate ($dH/dt$) is constant, we can rewrite the above expression as a function of time:  
\begin{equation}
M_m(t)= At + Bt^{r}. \label{Eq2}
\end{equation}
Where now, $A$ and $B$ are proportional to $A'$ and $B'$, and $r$ is the same of Eq. \ref{Eq2H}.

Our approximate model considers that in the time-interval $t_1 \leq t \leq t_{H^{-}}$ the system is relaxing due to the positive $H_{max}$ applied before, but it is already under the effect of a linearly varying negative $H$. Therefore, the equation describing the $M_1$ stretch is
\begin{equation}
\begin{split}
M_{1}(t)& = \{M_{sp} + M_{0}e^{-\left[(t-t_{1})/t_{p}\right]^{n}}\} \\
& - \{A(t-t_{1}) + B(t-t_{1})^{r}\}. \label{Eq3} \\
\end{split}
\end{equation}
The first bracket corresponds to the SG-like phase's relaxation from the previously applied positive $H_{max}$, with the same parameters of Eq. \ref{Eq1}. The second bracket accounts for the influence of the immediately applied negative $H$. In order to evaluate the parameters of the second bracket, we kept fixed those obtained from the fitting of Eq. \ref{Eq1}.

\begin{figure}
\begin{center}
\includegraphics[width=0.5 \textwidth]{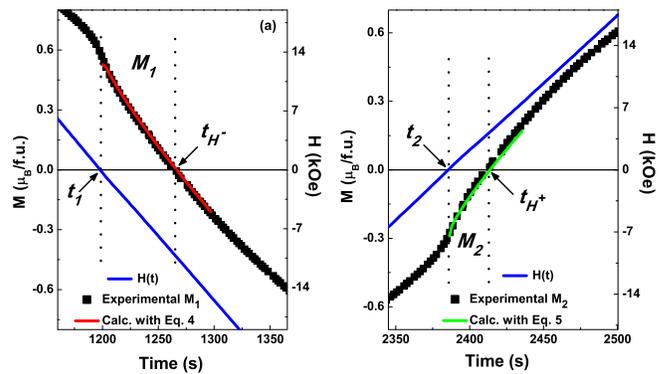}
\end{center}
\caption{Magnified view of the experimental and calculated $M_1$ (a) and $M_2$ (b) stretches of the $M(H)$ curve obtained for LSCMO at $T=5$ K and $|dH/dt|=150$ Oe/s. The blue solid line is the magnetic field time dependence.}
\label{FigZOOM_Sr}
\end{figure}

Fig. \ref{FigZOOM_Sr}(a) shows the fitting of $M_1$ with Eq. \ref{Eq3} for the $M(H)$ curve of LSCMO with $|dH/dt| = 150$ Oe/s, where one can see a very good agreement between the fitted and experimental results. The parameters obtained from the fitting are displayed on Table \ref{T1}. As can be noted from Table \ref{T1}, $A$ is negligible for any $|dH/dt|$ of the LSCMO sample, indicating that around $H^-$ the FM contribution is much larger than the AFM one.

Now, with the parameters obtained for the FM, AFM and SG-like phases, the $M_2$ stretch can be calculated. Here our approximate assumption is that, when the negative field is applied, not all SG-like moments have been flipped toward negative direction, but half are still relaxing from the positive field previously applied. Hence, the equation for $M_2$ becomes
\begin{equation}
\begin{split}
M_{2}(t)& = -\{M_{sp} + \frac{M_{0}}{2}e^{-\left[(t-t_{2})/t_{p}\right]^{n}}\} + \{\frac{M_{0}}{2}e^{-\left[(t-t_{1})/t_{p}\right]^{n}}\} \\
& + \{A(t-t_{2}) + B(t-t_{2})^{r}\}, \label{Eq4} \\
\end{split}
\end{equation}
where the first bracket represents the decay of the SG-like spins that are relaxing from the negative field applied before, the second one corresponds to the relaxation from the positive field previously applied, and the third represents the variation in the AFM/FM phases due to the just applied positive field. The decay of the SG-like moments pointing toward positive direction starts at $t_1$, while the decay of those pointing toward negative directions starts at $t_2$. Also, the FM phase spontaneous magnetization $M_{sp}$ is assumed to have been flipped toward negative direction due to the $H = -90$ kOe previously applied. Using the parameters obtained from Eqs. \ref{Eq1} and \ref{Eq2}, the $M_2$ stretch can be calculated. As can be seen in Fig. \ref{FigZOOM_Sr}(b), the curve obtained from the model is very similar to the experimental result. From the calculation of $M_2$ with Eq. \ref{Eq4} one gets $t^{calc}_{H^{+}}$ = 2412.5 s, yielding $H^{+}_{calc}=150\times(t^{calc}_{H^{+}}-t_{2})=4095$ Oe. Combining it to $H^{-}=-10305$ Oe obtained from the $M_1$ stretch, one gets $H_{EB}^{calc}=3105$ Oe, which is close to the experimentally observed value $H_{EB}^{exp}=3127$ Oe. It must be stressed that the theoretical curve of Fig. \ref{FigZOOM_Sr}(b) is not a fitting to the experimental $M_2$, but a calculated curve for which the parameters obtained from Eqs. \ref{Eq1} and \ref{Eq3} were kept fixed.

Once again, it must be emphasized that the above described model is an approximation that does not intend to perfectly describe the microscopic mechanism responsible for the complex magnetic behavior of the investigated systems, but only to shed some light on the ZEB effect and evidence the role played by the SG-like phase. There are several possible sources of discrepancies of our model to the real system, many of them being related to dynamics of the SG-like phase itself, but also to instrumental sources of imprecision (see SM \cite{SM}). Nevertheless, despite the above mentioned sources of imprecision, the proximity between the calculated and experimental results is remarkable.

A significant evidence of the validity of our model is that, according to it, $H_{EB}$ is expected to vary for different $|dH/dt|$. One can observe on Table \ref{T1} that both parameters of Eq. \ref{Eq1}, the $n$ exponent and $t_{p}$, change with $|dH/dt|$. These parameters govern the relaxation rate of the SG-like phase. For smaller/larger $n$, slower/faster will relax the magnetization \cite{Mydosh}, which increases/decreases both $|H^{-}|$ and $H^{+}$ in our model. However, since in Eq. \ref{Eq4} the two terms containing $n$ have opposites signs, the increase/decrease is always larger on $|H^{-}|$ than on $H^{+}$. Therefore, decreasing/increasing $n$ will lead to an enhancement/diminishment of $H_{EB}$. Although, the $H_{EB}$ of LSCMO and LCCMO has opposite trends in respect to $dH/dt$, our model captures the interplay between SG-like magnetization relaxation and exchange bias.

We can understand the observed values of $H_{EB}$ as a counterbalance between $|dH/dt|$, the relaxation of the SG-like phase, and the magnetic history of the $M(H)$ loop. We can think that the magnetic history of the magnetization loop is imprinted in the $M_{SG}$. So, a slow/fast SG-like relaxation corresponds to keeping an old/fresh memory of the magnetization loop history. The magnetic history of the $M_1$ stretch, and therefore of $H^{-}$, is quite different from those of $M_2$ and $H^{+}$. Depending on how old/fresh is the memory, larger/smaller will be the loop asymmetry, implying in larger/smaller $H_{EB}$ values. Since both $t_p$ and $n$ indicate that the magnetization relaxation is slower for LSCMO than for LCCMO, this corresponds to an older memory, the $H_{EB}$ for the first should be larger than for the latter.

\begin{figure}
\begin{center}
\includegraphics[width=0.45 \textwidth]{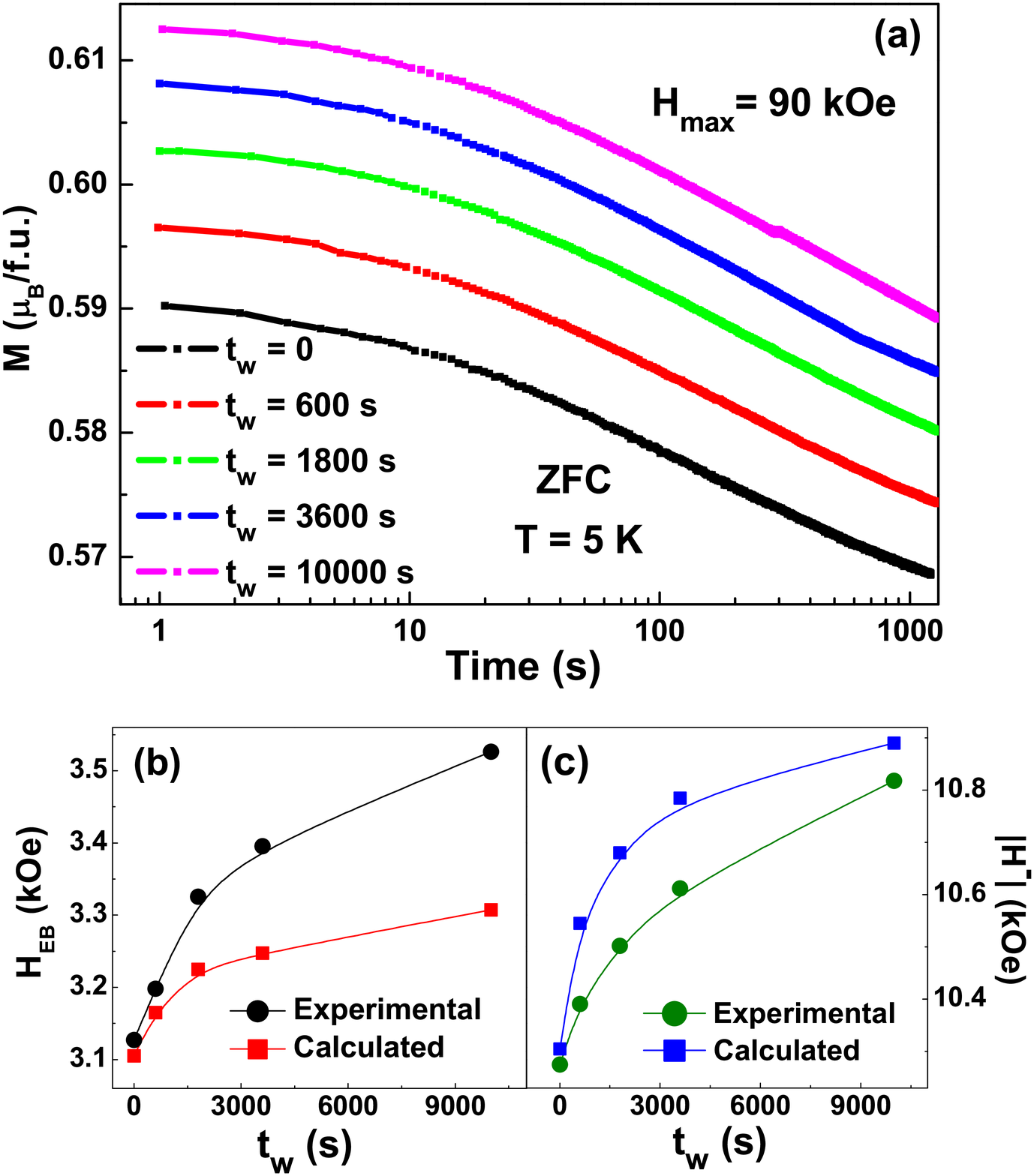}
\end{center}
\caption{(a) $IRM(t)$ curves for LSCMO at $T=5$ K and $|dH/dt|=150$ Oe/s, for different $t_w$ at $H_{max}=90$ kOe. (b) and (c) shows the evolution of $H_{EB}$ and $|H^{-}|$ as a function of $t_w$, respectively. The solid lines are guides for the eye.}
\label{FigIRM_tw9T_Sr}
\end{figure}

Our ZEB phenomenological model can also describe the change in $H_{EB}$ observed for the measurements where $M(H)$ loops were paused for a given interval $t_w$ at $H_{max}=90$ kOe. In Fig. \ref{FigIRM_tw9T_Sr}(a) are shown the $IRM(t)$ curves obtained for LSCMO for different $t_w$. Clearly, the magnetization increases as $t_w$ increases, leading to the increment of $M_0$ and $M_{sp}$ (see SM \cite{SM}). It can also be noted that the curve's slope does not change significantly for different $t_w$, since it affects mainly the amount of SG-like phase, while the parameters of Eq. \ref{Eq2} remain nearly the same. In this case, the $M_1$ and $M_2$ stretches are also calculated using Eqs. \ref{Eq3} and \ref{Eq4}, as done for $M(H)$ loops for $t_w = 0$, but keeping $A$, $B$ and $r$ parameters fixed with the values obtained for $t_w$ = 0. Figs. \ref{FigIRM_tw9T_Sr}(b) and (c) show the evolution of experimental and calculated $H_{EB}$ and $|H^{-}|$ with increasing $t_w$. It is worth noting that the $M_1$ stretch is no longer a fitting, but a calculation, which checks the model strength and allows a direct observation of the influence of the SG-like phase on the EB effect. Although there are quantitative differences between the experimental and calculated values, our model captures the  $H_{EB}$ trend observed. For LCCMO, the same overall behavior was found \cite{SM}.

For experiments pausing at $H = 0$ with different $t_w$, all parameters were kept fixed at the values obtained from the $t_w = 0$ curve. The $M_1$ and $M_2$ stretches were calculated from modified Eqs. \ref{Eq3} and \ref{Eq4}, in which we take into account the effect of $t_w$ on the SG-like phase, yielding:
\begin{equation}
\begin{split}
M_{1}(t)& = \{M_{sp} + M_{0}e^{-\left[(t-t_{1})/t_{p}\right]^{n}}\} \\
& - \{A[t-(t_{1}+t_w)] + B[t-(t_{1}+t_w)]^{r}\}, \label{Eq5} \\
\end{split}
\end{equation}
\begin{equation}
\begin{split}
M_{2}(t)& = -\{M_{sp} + \frac{M_{0}}{2}e^{-\left[(t-t_{2})/t_{p}\right]^{n}}\} + \{\frac{M_{0}}{2}e^{-\left[(t-t_{1})/t_{p}\right]^{n}}\} \\
& + \{A[t-(t_{2}+t_w)] + B[t-(t_{2}+t_w)]^{r}\}. \label{Eq6} \\
\end{split}
\end{equation}
The $M_1$ and $M_2$ curves calculated with Eqs. \ref{Eq5} and \ref{Eq6} agree very well to the experimental results, as can be seen on Fig. \ref{Fig_Wait0T_Sr}(a).

\begin{figure}
\begin{center}
\includegraphics[width=0.45 \textwidth]{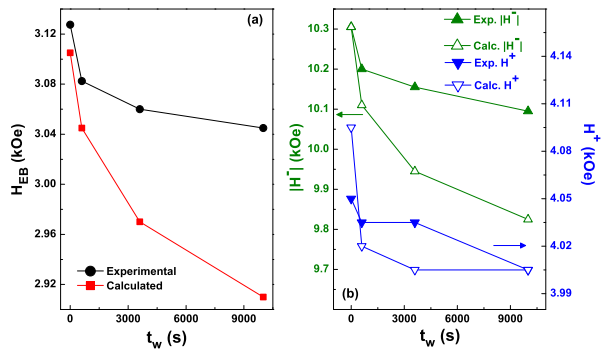}
\end{center}
\caption{Experimental and calculated (a) $H_{EB}$ and (b) $H^{+}$ and $|H^{-}|$ as a function of $t_w$ obtained from the $M(H)$ loops of LSCMO in which the experiment was paused at $H = 0$. The solid lines are guides for the eye.}
\label{Fig_wt0T_Heb}
\end{figure}

In this case, we observed a decrease of $H_{EB}$ as a function of $t_w$, related to the reduction of both $|H^{-}|$ and $H^{+}$, see Figs. \ref{Fig_wt0T_Heb}(a) and (b). This can be understood as a partial refresh of the memory imprinted on the $M_{SG}$ by the magnetic history, when the loop is paused at $H = 0$. In both descending and ascending branches, during the time-interval $t_w$ the SG-like magnetization relaxes exponentially toward zero. Once the exchange bias is due to the coupling between the FM and SG-like phases, and considering that the latter decreases, it is expected that $H_{EB}$ will decrease as well. In the limiting situation, pausing at $H=0$ for an infinite time, or alternatively, for a very fast $M_{SG}$ relaxation, the memory will be completely erased, and $H_{EB}$ will be zero.

\section{Summary}

In summary, we have investigated in detail the dynamics of magnetization for two RSG compounds that are representative ZEB materials: La$_{1.5}$Sr$_{0.5}$CoMnO$_{6}$ and La$_{1.5}$Ca$_{0.5}$CoMnO$_{6}$. Different experiments were carried out to elucidate how the SG-like phase influences the ZEB effect. For comparison, the magnetic properties of the non-RSG and non-ZEB La$_{2}$CoMnO$_{6}$ compound were also investigated. We found strong evidences that the unusual magnetic relaxation of the glassy moments is strongly correlated to the loop-shifts. We also propose a simple phenomenological model, based on the fractionated flipping of the SG-like spins and on their unusual dynamics, to describe the ZEB effect. The calculated results agree with the experimental values. Based on the model, the large difference between the ZEB effect observed for La$_{1.5}$Sr$_{0.5}$CoMnO$_{6}$ and La$_{1.5}$Ca$_{0.5}$CoMnO$_{6}$ could be explained in terms of magnetization relaxation of the SG-like phase. Although it must be checked for other ZEB materials, the model indicates that the dynamics of the SG-like phases present on these RSG systems is key to the emergence of ZEB effect.

\begin{acknowledgements}
We thank Prof. A. P. Guimar\~aes for critical reading of the manuscript. This work was supported by Conselho Nacional de Desenvlovimento Cient\'{i}fico e Tecnol\'{o}gico (CNPq) [No. 100134/2016-0], Funda\c{c}\~{a}o Carlos Chagas Filho de Amparo \`{a} Pesquisa do Estado do Rio de Janeiro (FAPERJ) and Coordena\c{c}\~{a}o de Aperfei\c{c}oamento de Pessoal de N\'{i}vel Superior (CAPES).
\end{acknowledgements}


\begin{thebibliography}{99}

\bibitem{Fischer} K. H. Fischer and J. A. Hertz, \textit{Spin Glasses} (Cambridge University Press, Cambridge, 1991).

\bibitem{Stein} D. L. Stein and C. M. Newman, \textit{Spin Glasses and Complexity} (Princeton University Press, Princeton, 2013).

\bibitem{Mydosh} J. A. Mydosh, \textit{Spin Glasses: An Experimental Introduction} (Taylor \& Francis, London, 1993).

\bibitem{Nogues} J. Nogu\'{e}s and I. K. Schuller, J. Magn. Magn. Mater. \textbf{192}, 203 (1999).

\bibitem{Dagotto} E. Dagotto, Science \textbf{318}, 1076 (2007).

\bibitem{Wang} B. M. Wang, Y. Liu, P. Ren, B. Xia, K. B. Ruan, J. B. Yi, J. Ding, X. G. Li, and L. Wang, Phys. Rev. Lett. \textbf{106}, 077203 (2011).

\bibitem{Maity} T. Maity, S. Goswami, D. Bhattacharya, and S. Roy, Phys. Rev. Lett. \textbf{110}, 107201 (2013).

\bibitem{Nayak} A. K. Nayak, M. Nicklas, S. Chadov, C. Shekhar, Y. Skourski, J. Winterlik, and C. Felser, Phys. Rev. Lett. \textbf{110}, 127204 (2013).

\bibitem{Tian} F. Tian, K. Cao, Y. Zhang, Y. Zeng, R. Zhang, T. Chang, C. Zhou, M. Xu, X. Song, and S. Yang, Sci. Rep. \textbf{6}, 30801 (2016).

\bibitem{Vasala} S. Vasala and M. Karpinnen, Prog. Solid State Chem. \textbf{43}, 1 (2015).

\bibitem{Serrate} D. Serrate, J. M. De Teresa, and M. R. Ibarra, J. Phys.: Condens. Matter \textbf{19}, 023201 (2007).

\bibitem{MeuPRB} L. T. Coutrim, E. M. Bittar, F. Stavale, F. Garcia, E. Baggio-Saitovitch, M. Abbate, R. J. O. Mossanek, H. P. Martins, D. Tobia, P. G. Pagliuso, and L. Bufai\c{c}al, Phys. Rev. B \textbf{93}, 174406 (2016).

\bibitem{Murthy} J. Krishna Murthy and A. Venimadhav, Appl. Phys. Lett. \textbf{103}, 25410 (2013).

\bibitem{MeuJMMM} L. Bufaical, R. Finkler, L. T. Coutrim, P. G. Pagliuso, C. Grossi, F. Stavale, E. Baggio-Saitovitch, and E. M. Bittar, J. Magn. Magn. Mater. \textbf{433}, 271 (2017).

\bibitem{Huang} S. Huang, L. R. Shi, Z. M. Tian, H. G. Sun, and S. L. Yuan, J. Magn. Magn. Mater. \textbf{394}, 77 (2015).

\bibitem{Xie} L. Xie and H. G. Zhang, Curr. Appl. Phys. \textbf{18}, 261 (2018).

\bibitem{Singh} M. P. Singh, K. D. Truong, and P. Fournier, Appl. Phys. Lett. \textbf{91}, 042504 (2007).

\bibitem{Burnus} T. Burnus, Z. Hu, H. H. Hsieh, V. L. J. Joly, P. A. Joy, M. W. Haverkort, Hua Wu, A. Tanaka, H.-J. Lin, C. T. Chen, and L. H. Tjeng, Phys. Rev. B \textbf{77}, 125124 (2008).

\bibitem{Mir} L. L\'{o}pez-Mir, R. Galceran, J. Herrero-Mart\'{i}n, B. Bozzo, J. Cisneros-Fern\'{a}ndez, E. V. Pannunzio Miner, A. Pomar, L. Balcells, B. Mart\'{i}nez, and C. Frontera, Phys. Rev. B \textbf{95}, 224434(2017).

\bibitem{SM} See Supplemental Material at ??? for details of the crystals growth, magnetization measurements and calculations with the proposed equations. It includes Refs. 21-23.

\bibitem{GSAS} A.C. Larson, R.B. Von Dreele, General Structure Analysis System GSAS, Los Alamos National Laboratory Report (2001).

\bibitem{Murthy2} J. Krishna Murthy, K. D. Chandrasekhar, H. C. Wu, H. D. Yang, J. Y. Lin, and A. Venimadhav, J. Phys.: Condens. Matter. \textbf{28},086003 (2016).

\bibitem{Frey} M. H. Frey and D. A. Payne, Phys. Rev. B \textbf{54}, 3158 (1996).

\bibitem{Gokemeijer} N. J. G\"{o}kemeijer and C. L. Chien, J. Appl. Phys. \textbf{85}, 5516 (1999).

\bibitem{IRM} Conventionally, to measure the remnant magnetization of a SG material as a function of $t$, the system is cooled in the presence of a small $H$, which is supressed at $T<T_{sg}$ and then $M$ is computed as a function of $t$. Since our propose here is to model ZFC $M(H)$ loops, the following protocol was adopted to measure $IRM(t)$ curves: firstly the system was ZFC down to 5 K. Then $H$ was increased up to 90 kOe and subsequently decreased down to zero. After this protocol,$M$ was measured as a function of $t$. Note that this protocol, carried prior to each $IRM(t)$ measurement, corresponds to the virgin curve and the descending branch of a $M(H)$ loop.

\bibitem{Manna} K. Manna, R. S. Joshi, S. Elizabeth, and P. S. Kumar, Appl. Phys. Lett. \textbf{104}, 202905 (2015).

\bibitem{Murthy3} J. Krishna Murthy, K. D. Chandrasekhar, S. Murugavel, and A. Venimadhav, J. Mater. Chem. C \textbf{3},836 (2015).

\bibitem{Trutt} F. C. Trutt, E. A. Erd\'{e}lyi, and R. E. Hopkins, IEEE Trans. on Powder Apparatus and Systems \textbf{PAS-87}, 665 (1968).

\end{thebibliography}
\end{document}